\documentclass[letterpaper, 10 pt, conference]{ieeeconf}
\IEEEoverridecommandlockouts
\usepackage[utf8]{inputenc}
\usepackage{amsmath}
\usepackage{amssymb}
\usepackage{graphicx,balance}
\usepackage{pifont}
\usepackage{bm}
\usepackage{dsfont}
\usepackage{subcaption}
\usepackage{tikz}

\usepackage{siunitx}

\usepackage{amsfonts}
\usepackage{url}
\usepackage[pdftex,plainpages = false, colorlinks=true, linkcolor=black, citecolor = black, urlcolor = blue,pagebackref=false,hypertexnames=false, plainpages=false, pdfpagelabels]{hyperref}
\usepackage{balance}
\usepackage[capitalize]{cleveref}
\usepackage[sort,compress]{cite}

\newtheorem{theorem}{Theorem}[section]

\crefname{section}{Section}{Sections}
\crefname{theorem}{Theorem}{Theorems}
\crefname{lemma}{Lemma}{Lemmas}
\crefname{table}{Table}{Tables}
\crefformat{equation}{(#2#1#3)}
\crefname{algocf}{Algorithm}{Algorithms}
\Crefname{algocf}{Algorithm}{Algorithms}
\crefname{ALC@unique}{Line}{Lines}

\newcommand{\x}{\mathbf{x}}

\newcommand{\R}{\mathds{R}}

\newcommand{\istate}{k}

\newcommand{\CROWNAu}{\mathbf{\Psi}}
\newcommand{\CROWNAl}{\mathbf{\Phi}}
\newcommand{\CROWNbu}{\bm{\alpha}}
\newcommand{\CROWNbl}{\bm{\beta}}

\usepackage{accents}
\newcommand{\ubar}[1]{\underaccent{\bar}{#1}}

\newcommand{\hybrid}{HyBReach-LP\textsuperscript{+}}
\usepackage{paralist}

\usepackage[addedmarkup=bf]{changes}
\definechangesauthor[name={MFE}, color={blue}]{MFE}
\definechangesauthor[name={JH}, color={red}]{JH}
\definechangesauthor[name={NR}, color={orange}]{NR}

\newcommand\Mark[1]{\textsuperscript#1}

\crefformat{chapter}{\S#2#1#3}
\crefmultiformat{chapter}{\S\S#2#1#3}{and~#2#1#3}{, #2#1#3}{, and~#2#1#3}
\crefformat{section}{\S#2#1#3}
\crefmultiformat{section}{\S\S#2#1#3}{and~#2#1#3}{, #2#1#3}{, and~#2#1#3}

\usepackage{algorithm,algorithmic}

\title{\LARGE \bf
A Hybrid Partitioning Strategy for Backward Reachability of Neural Feedback Loops
}

\author{Nicholas Rober\Mark{1}, Michael Everett\Mark{1}, Songan Zhang\Mark{2}, and Jonathan P.\ How\Mark{1} %
\thanks{\Mark{1}Aerospace Controls Laboratory, Massachusetts Institute of Technology, Cambridge, MA, USA. e-mail: {\tt \small \{nrober,mfe,jhow\}@mit.edu}. Research supported by Ford Motor Company.}
\thanks{\Mark{2}Ford Motor Company, Dearborn, MI, USA. e-mail: {\tt \small szhang117@ford.com}.}}

\begin{document}

\maketitle

\begin{abstract}
As neural networks become more integrated into the systems that we depend on for transportation, medicine, and security, it becomes increasingly important that we develop methods to analyze their behavior to ensure that they are safe to use within these contexts.
The methods used in this paper seek to certify safety for closed-loop systems with neural network controllers, i.e., neural feedback loops, using backward reachability analysis.
Namely, we calculate backprojection (BP) set over-approximations (BPOAs), i.e., sets of states that lead to a given target set that bounds dangerous regions of the state space. 
The system's safety can then be certified by checking its current state against the BPOAs. 
While over-approximating BPs is significantly faster than calculating exact BP sets, solving the relaxed problem leads to conservativeness.
To combat conservativeness, partitioning strategies can be used to split the problem into a set of sub-problems, each less conservative than the unpartitioned problem.
We introduce a hybrid partitioning method that uses both target set partitioning (TSP) and backreachable set partitioning (BRSP) to overcome a lower bound on estimation error that is present when using BRSP.
Numerical results demonstrate a near order-of-magnitude reduction in estimation error compared to BRSP or TSP given the same computation time.

%
\end{abstract}


\section{Introduction}

Neural networks (NNs) are important tools for robotic systems due to their ability to capture complex relationships between different signals.
Increasingly sophisticated training methods and architectures have broadened the range of applications to which NNs can be applied, even expanding their use to safety-critical applications such as autonomous driving \cite{zhang2019discretionary} and aircraft collision avoidance \cite{julian2019deep}.
While these NNs can be empirically demonstrated as safe in nominal scenarios, there remains a need to provide formal statements about their safety before they can be applied to safety-critical applications where adversarial examples and attacks \cite{kurakin2016adversarial, yuan2019adversarial, madry2017towards} can cause dangerous situations.

\begin{figure}[t]
\centering
    \includegraphics[width=0.95\columnwidth]{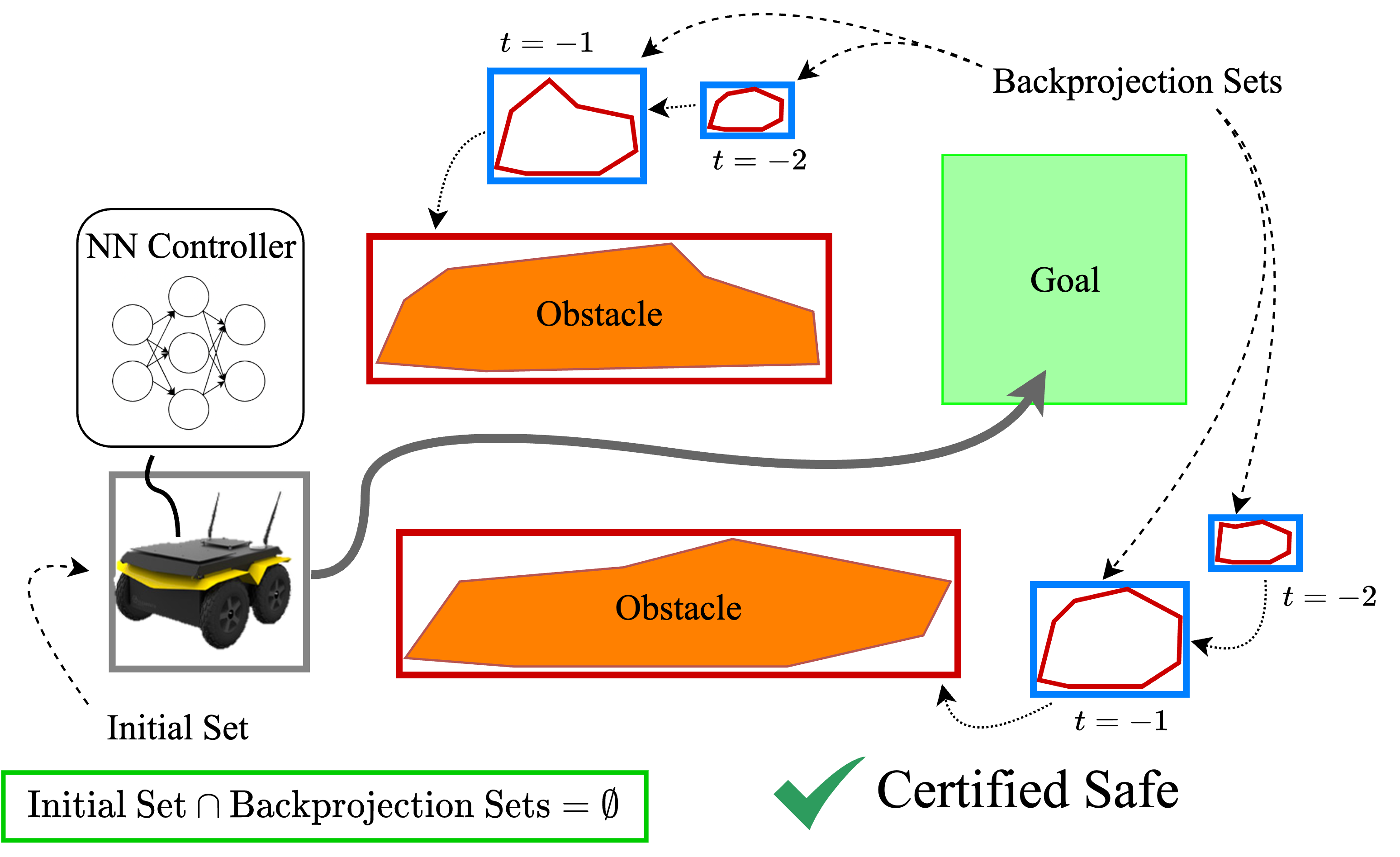}
    \caption{Backward reachability strategy for collision avoidance. The backprojection set estimates (blue) define the set of states that will cause the system to collide with the target sets (red) within a 2 second time horizon.}
    \label{fig:intro:backward_reach}
\vspace{-5mm}
\end{figure}

To this end, numerous tools \cite{zhang2018efficient, weng2018towards, xu2020automatic, raghunathan2018semidefinite, tjeng2017evaluating, katz2019marabou, katz2017reluplex, jia2021verifying} have been developed to conduct open-loop NN analysis allowing users to certify properties about NNs in isolation.
For neural feedback loops (NFLs), which are closed-loop systems with NN control policies, analysis is typically taken one step further by analyzing the entire closed-loop system using reachability analysis \cite{dutta2019reachability, huang2019reachnn, ivanov2019verisig, fan2020reachnn, xiang2020reachable, hu2020reach, sidrane2021overt, everett2021reachability, vincent2021reachable, bak2022closed, rober2022backward}.
Reachability analysis of NFLs can generally be split into two groups: Forward reachability analysis \cite{dutta2019reachability, huang2019reachnn, ivanov2019verisig, fan2020reachnn, xiang2020reachable, hu2020reach, sidrane2021overt, everett2021reachability, vincent2021reachable} determines the set of states the system could reach given a set of possible initial states.
Conversely, backward reachability analysis \cite{vincent2021reachable, everett2021reachability, bak2022closed, rober2022backward}, considered in this paper, determines the set of possible previous states, referred to as backprojection (BP) sets, given a set of final states, called the target set.
An example of using backward reachability analysis to certify safety in an obstacle avoidance scenario with a ground vehicle and two obstacles is shown in \cref{fig:intro:backward_reach}.

Due to the computational expense associated with calculating exact BP sets, previous work by the authors \cite{rober2022backward} introduced a technique to calculate BP set over-approximations (BPOAs) to capture all states that reach the target set.
The calculation of BPOAs is dependent on NN relaxation techniques such as \cite{zhang2018efficient} that can efficiently find outer bounds on a NN's output.
While efficient, relaxing the NN in this way inherently introduces some conservativeness because the relaxation bounds a potentially highly-nonlinear NN with lower-order, e.g., linear or quadratic, bounds.
To combat conservativeness, previous work \cite{xiang2018output, wang2018formal, rubies2019fast, everett2020robustness, rober2022partition} investigated partitioning strategies that solve for BP sets of subsets of the target set, resulting in less conservativeness than the solution of the original problem because relaxations are applied over a smaller domain.

Ref.~\cite{rober2022partition} investigated backreachable set partitioning (BRSP) wherein the set of all possible previous states, i.e., the backreachable (BR) set, is partitioned to get a set of tightened NN relaxations that can be used to calculate BPOAs.
However, it was shown that this technique led to a lower bound on the approximation error that could not be improved by more BRSP.
This work introduces the concept of target set partitioning (TSP), and develops \hybrid{}: a hybrid partitioning algorithm employing both BRSP and TSP to reduce approximation error by an order of magnitude when compared with \cite{rober2022partition} and overcome the error bound.
In summary, the contributions of this work include the development of \hybrid{}, a hybrid partitioning strategy for backward reachability analysis. 
Specifically, we combine ideas from \cite{rober2022partition} with partitioning concepts from forward reachability to overcome a performance limitation of prior methods. 
Finally, we showcase numerical results demonstrating a near-order-of-magnitude improvement in approximation error from previous techniques given the same computation time.

\section{Preliminaries}

\subsection{NFL System Dynamics}
We consider the dynamics given by the linear time-invariant system
\begin{align}
\begin{split}
    \mathbf{x}_{t+1} & = \mathbf{A} \mathbf{x}_{t} + \mathbf{B} \mathbf{u}_t + \mathbf{c} \triangleq f(\mathbf{x}_{t}, \mathbf{u}_t) \label{eqn:lti_dynamics}
\end{split}
\end{align}
where $\mathbf{x}_{t} \in \mathds{R}^{n_x}$ is the system's state, $\mathbf{u}_t \in \mathds{R}^{n_u}$ is the input, $\mathbf{A} \in \mathds{R}^{n_x \times n_x}$ and $\mathbf{B} \in \mathds{R}^{n_x \times n_u}$ are known system matrices, and $\mathbf{c} \in \mathds{R}^{n_x}$ is a constant known exogenous input. We also assume $\mathbf{x_t} \in \mathcal{X}$ and $\mathbf{u_t} \in \mathcal{U}$ where $\mathcal{X}$ and $\mathcal{U}$ are convex sets defining the operating region of the state space and the control limits, respectively.
We can then write the closed-loop NFL formally as 
\begin{align}
\begin{split}
    \mathbf{x}_{t+1} & = \mathbf{A} \mathbf{x}_{t} + \mathbf{B}\pi(\mathbf{x}_t) + \mathbf{c} \triangleq p(\mathbf{x}_t;\pi)
    \label{eqn:nfl}
\end{split}
\end{align}
where $\pi(\cdot)$ is a state-feedback control policy comprised of a feedforward NN as discussed in the following section.
Notice that $p$ is explicitly only a function of $\mathbf{x}_t$ with an implicit dependence on the state feedback from $\pi$.


\subsection{Control Policy Structure}

Denote a feedforward NN with $K$ hidden layers as $\pi(\cdot)$ where each layer $k \in [K+1]$ has $n_k$ neurons with the notation $[j]$ representing the set $\{0,1,\ldots,j\}$.
We introduce the $k^\mathrm{th}$ layer's weight matrix $\mathbf{W}^{(k)}\in\R^{n_{k+1}\times n_{k}}$, bias vector $\mathbf{b}^{(k)}\in\R^{n_{k+1}}$, and coordinate-wise activation function $\sigma^{(k)}(\cdot): \R^{n_{k+1}} \to \R^{n_{k+1}}$. The activation functions can be any combination of tanh, sigmoid, ReLU (i.e., $\sigma(\mathbf{z})=\mathrm{max}(0,\mathbf{z})$) and others considered by \cite{zhang2018efficient}.
Given an input $\mathbf{x} \in \mathds{R}^{n_0}$, the output of the NN, i.e., $\pi(\mathbf{x})$ is calculated using 
\begin{align*}
\begin{split}
    & \x^{(0)} = \x \\
    & \x^{(k+1)} = \sigma^{(k)}(\mathbf{W}^{(k)} \x^{(k)}+\mathbf{b}^{(k)}),\ \forall k\in[K-1] \\
    & \pi(\x) = \mathbf{W}^{(K)} \x^{(K)}+\mathbf{b}^{(K)}.
\end{split}
\end{align*}

\subsection{Neural Network Robustness Verification}
A key step in efficiently conducting reachability analysis for an NFL~\cref{eqn:nfl} is to relax the nonlinearities introduced by the NN's activation functions. 
Thus, denoting $\mathcal{B}(\mathbf{a}, \mathbf{b})$ as the hyper-rectangular set  $\{\mathbf{x} |\ \mathbf{a} \leq \mathbf{x} \leq \mathbf{b}\}$, where $\mathbf{a},\mathbf{b} \in \mathds{R}^{n_x}$ and $\mathbf{a} < \mathbf{b}$, we use the following theorem to obtain affine bounds on the NN output.
\begin{theorem}[\!\!\cite{zhang2018efficient}, Convex Relaxation of NN]\label{thm:crown_particular_x}
Given an $m$-layer neural network control policy $\pi:\R^{n_x}\to\R^{n_u}$ and a hyper-rectangular region $\mathcal{B}(\mathbf{a},\mathbf{b})$, there exist two explicit functions $\pi_j^L: \R^{n_x}\to\R^{n_u}$ and $\pi_j^U: \R^{n_x}\to\R^{n_u}$ such that $\forall j\in [n_m], \forall \mathbf{x}\in\mathcal{B}(\mathbf{a},\mathbf{b})$, the inequality $\pi_j^L(\mathbf{x})\leq \pi_j(\mathbf{x})\leq \pi_j^U(\mathbf{x})$ holds true, where
\begin{equation*}
\label{eq:f_j_UL}
    \pi_{j}^{U}(\x) = \CROWNAu_{j,:} \x + \CROWNbu_j, \quad
    \pi_{j}^{L}(\x) = \CROWNAl_{j,:} \x + \CROWNbl_j,
\end{equation*}
where $\CROWNAu, \CROWNAl \in \R^{n_u \times n_x}$ and $\CROWNbu, \CROWNbl \in \R^{n_u}$ are defined recursively using NN weights, biases, and activations (e.g., ReLU, sigmoid, tanh), as detailed in~\cite{zhang2018efficient}.
\end{theorem}

\subsection{Backprojection Sets \& Time Notation}
The goal of this work is to over-approximate the set of states that reach a given target set $\mathcal{X}_T$ within $\tau$ timesteps, i.e., the BP sets for $t \in \mathcal{T} \triangleq \{-\tau, -\tau + 1, ... , -1\}$.
The first BP set can be written formally as 
\begin{equation}
    \mathcal{P}_{-1}(\mathcal{X}_T) = \{ \mathbf{x}\ \lvert\ p(\mathbf{x};\pi) \in \mathcal{X}_T \}.
    \label{eqn:one_step_backprojection_sets}
\end{equation}
More generally, we can define the BP set at time $t$ as
\begin{equation}
    \mathcal{P}_{t}(\mathcal{X}_T) = \{ \mathbf{x}\ \lvert\ p^{(-t)}(\mathbf{x}; \pi) \in \mathcal{X}_T \} 
    \label{eqn:backprojection_sets}
\end{equation}
where $t<0$ and the notation $p^n$ represents the $n^{\mathrm{th}}$ iterate of $p$, i.e. $p^{n+1} \triangleq p \circ p^{n}$.
From here we will omit the argument $\mathcal{X}_T$, i.e., simply writing $\mathcal{P}_{t}$, unless necessary for clarity.

\begin{figure}[h]
\setlength\belowcaptionskip{-0.7\baselineskip}
\centering
\captionsetup[subfigure]{aboveskip=-1pt,belowskip=-1pt}
    \begin{tikzpicture}[fill=white]
        \node[anchor=south west,inner sep=0] (image) at (0,0) {\includegraphics[width=\columnwidth]{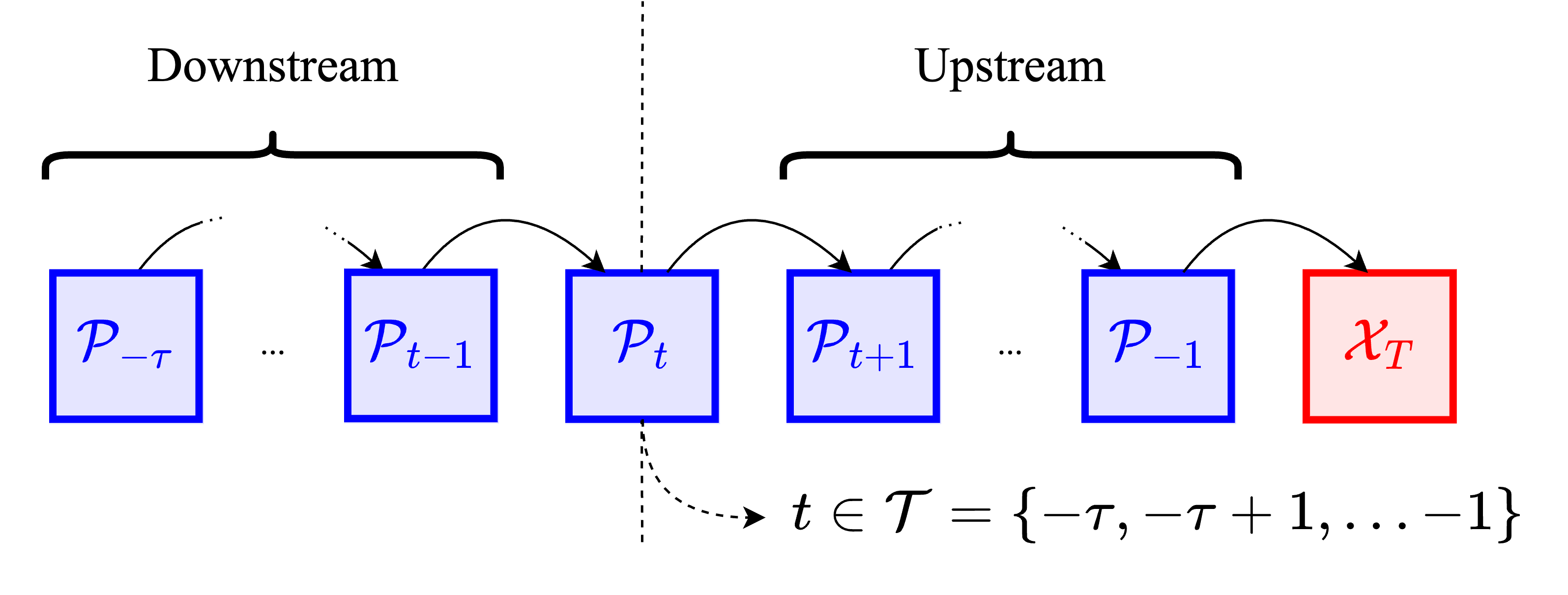}};
    \end{tikzpicture}
    \caption{Convention for time-indexing of sets.}
    \label{fig:prelim:time}
\end{figure}

A visual explanation of the time-indexing convention is given in \cref{fig:prelim:time}.
Notice that the values of $t$ are negative, so increasing $t$ represents the system's progression towards the target set.
For brevity of future discussion, when we consider a set at time $t$, the sets further into the time horizon (i.e., ${\mathcal{P}}_{t-1}, ..., {\mathcal{P}}_{-\tau}$) are referred to as \textit{downsteam}, while those that are fewer steps from $\mathcal{X}_T$ (i.e., ${\mathcal{P}}_{t+1}, ..., {\mathcal{P}}_{-1}$) are referred to as \textit{upstream}.

\subsection{Backreachable Sets \& BPOAs}
To use forward NN analysis tools such as \cite{zhang2018efficient}, \cref{thm:crown_particular_x} requires that there must be a region over which the NN bounds can be calculated.
Thus, the procedure to calculate a BPOA introduced in \cite{rober2022backward} is to first find the set of all possible previous states $\bar{\mathcal{R}}_t$, relax the NN within $\bar{\mathcal{R}}_t$ to obtain the bounds $\pi^L_t(\x_t)$ and $\pi^U_t(\x_t)$, and use the bounds to calculate the BPOA $\bar{\mathcal{P}}_t$.
We therefore first define the set of all possible previous states, i.e., the BR set, as
\begin{equation}
    \bar{\mathcal{R}}_{t}(\bar{\mathcal{P}}_{t+1}) \triangleq \{ \mathbf{x}\ \lvert\ \exists\mathbf{u} \in \mathcal{U} \mathrm{\ s.t.\ }
     f(\mathbf{x},\mathbf{u}) \in \bar{\mathcal{P}}_{t+1} \},\
    \label{eqn:backreachable_sets}
\end{equation}
where $\bar{\mathcal{P}}_0 \triangleq \mathcal{X}_T$.
Finally, to provide an over-approximation of the BP set defined in \cref{eqn:one_step_backprojection_sets} a one-step BPOA can be defined using the NN relaxations from \cref{thm:crown_particular_x}, i.e.,
\begin{align}
    \bar{\mathcal{P}}_{-1} \triangleq \{ \mathbf{x}\ \lvert\ & \exists\mathbf{u} \in [\pi^L(\x), \pi^U(\x)] \mathrm{\ s.t.\ } \label{eqn:backprojection_set_over} \\ \nonumber
     &f(\mathbf{x},\mathbf{u}) \in \mathcal{X}_T \},\
\end{align}
We will describe the formation of $\bar{\mathcal{P}}_t$ (i.e., the BPOA at time $t$) more precisely in \cref{sec:approach}, but for now we point out that $\bar{\mathcal{P}}_t$ must lead to states that successively pass through the upstream BPOAs while satisfying the NN relaxations at each step and terminate in $\mathcal{X}_T$, i.e.,
\begin{equation}
\bar{\mathcal{P}}_t \! \triangleq \! \left\{
\! \mathbf{x}_{\mathfrak{t}} \left| \ 
\begin{aligned}
& \exists\mathbf{u}_t \in [\pi^L_t(\x_t), \pi^U_t(\x_t)] \mathrm{\ s.t.\ } \\
& f(\mathbf{x}_t,\mathbf{u}_t) = \mathbf{x}_{t+1} \in \bar{\mathcal{P}}_{t+1} \\
& \vdots \\
& \exists\mathbf{u}_{{-}1} \in [\pi^L_{{-}1}(\x_{{-}1}), \pi^U_{{-}1}(\x_{{-}1})] \mathrm{\ s.t.\ } \\
& f(\mathbf{x}_{{-}1},\mathbf{u}_{{-}1}) \in \mathcal{X}_T \\
\end{aligned} \; \; \right.\kern-\nulldelimiterspace 
\right\}.
\label{eqn:OBPAs}
\end{equation}

\subsection{Backreachable Set Partitioning} \label{sec:prelim:partitioning}

\begin{figure}[t]
\setlength\belowcaptionskip{-0.7\baselineskip}
\centering
\captionsetup[subfigure]{aboveskip=-1pt,belowskip=-1pt}
    \begin{tikzpicture}[fill=white]
        \node[anchor=south west,inner sep=0] (image) at (0,0) {\includegraphics[width=\columnwidth]{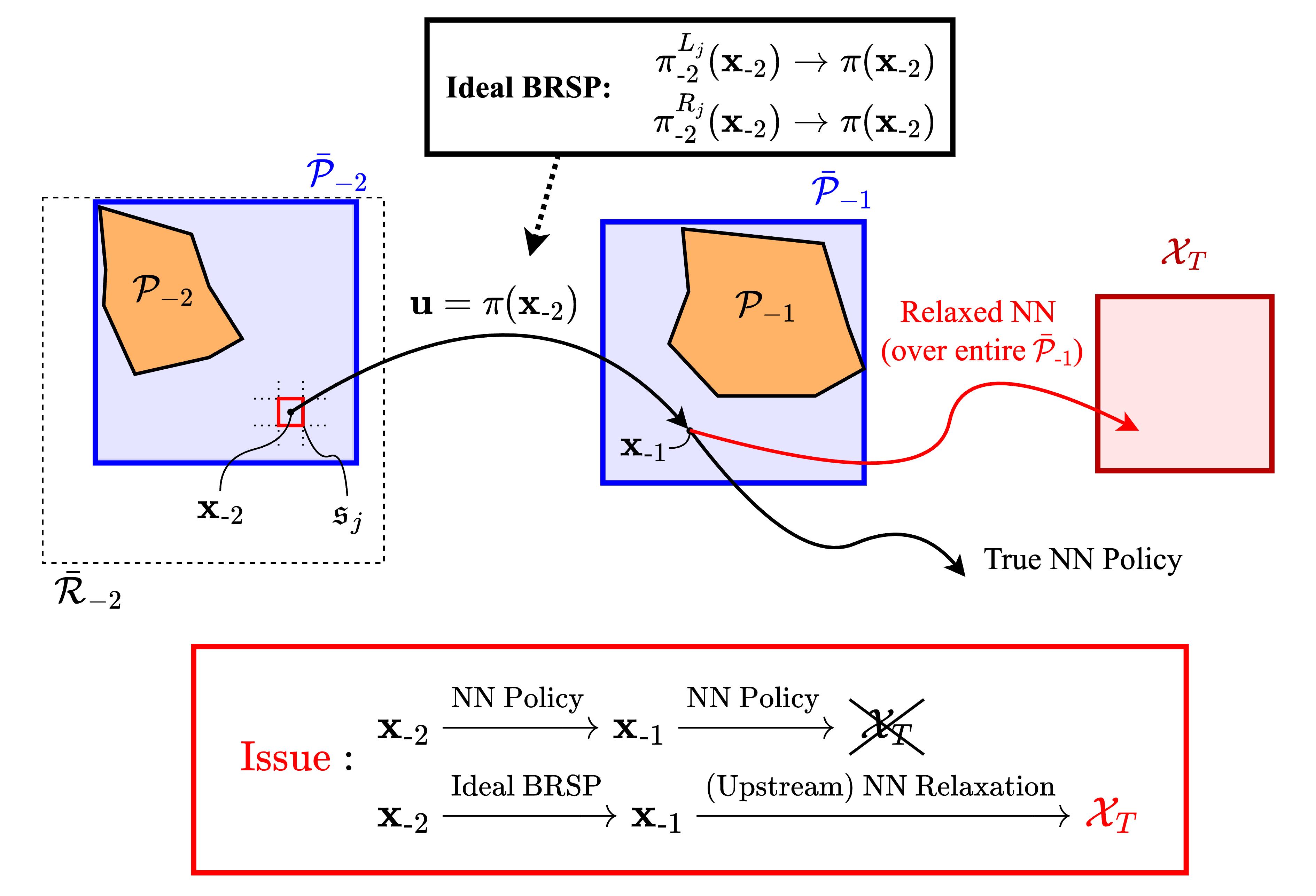}};
    \end{tikzpicture}
    \caption{Source of conservativeness that cannot be fixed with BRSP. To calculate $\bar{\mathcal{P}}_{-2}$, LPs require that the system reach $\mathcal{X}_T$ by traveling through $\bar{\mathcal{P}}_{-1}$ and \emph{satisfying a NN relaxation valid over all of $\bar{\mathcal{P}}_{-1}$}. 
    Thus, regardless of how well BRSP approximates $\pi$, any state in $\bar{\mathcal{P}}_{-2}$ that reaches $\mathbf{x}_{-1}$ can reach $\mathcal{X}_T$ given the upstream relaxation over $\bar{\mathcal{P}}_{-1}$, leading to conservativeness in $\bar{\mathcal{P}}_{-2}$.}
    \label{fig:prelim:pure_brsp_limit}
\end{figure}

To reduce conservativeness in  $\bar{\mathcal{P}}_t$, \cite{rober2022partition} investigated the strategy of backreachable set partitioning (BRSP) wherein $\bar{\mathcal{R}}_t$ is split into a set of partitions $\mathcal{S}$ and $\pi$ is relaxed over these smaller partitions to get relatively tight bounds for each BR set partition $\mathfrak{s}_j \in \mathcal{S}$, i.e., $\pi^{L_j}_{t}(\mathbf{x})$ and $\pi^{U_j}_{t}(\mathbf{x})$. 
While this enabled efficient computation of BPOAs with reduced conservativeness, \cref{fig:prelim:pure_brsp_limit} illustrates a limitation of this strategy.
When $\bar{\mathcal{P}}_{-1}$ is calculated, BRSP is used to generate tightened bounds on the NN. 
However, when calculating $\bar{\mathcal{P}}_{-2}$ with LPs, the partition-specific bounds originally calculated for time $t=-1$ must be replaced with bounds that are valid for $\bar{\mathcal{P}}_{-1}$ in its entirety. 
This idea extends to all $\bar{\mathcal{P}}_{t}$: BRSP provides tightened bounds for the current BPOA, but cannot reduce conservativeness associated with the relaxations of the upstream BPOAs.
The issue highlighted in \cref{fig:prelim:pure_brsp_limit} shows that to reduce conservativeness beyond what is possible using BRSP, the relaxations associated with the upstream BPOAs must be improved.

\section{Approach}
\label{sec:approach}
From \cref{sec:prelim:partitioning}, it is apparent that one way of improving the performance of \cite{rober2022partition} is by finding a way to maintain the tightened bounds calculated at time $t$ for the calculation of the downstream BPOAs.
With this in mind, this section first introduces target set partitioning (TSP) and contrasts it with BRSP.
Specifically, we explain how BRSP provides tightened NN relaxations for time $t$ while TSP allows the relaxations of the upstream BPOAs to remain tightened.
Finally, we improve \cite{rober2022partition} by leveraging TSP, resulting in \hybrid{}, an algorithm for calculating BPOAs using a hybrid-partitioning strategy.

\begin{figure}[t]
\setlength\belowcaptionskip{-0.7\baselineskip}
\centering
\captionsetup[subfigure]{aboveskip=-1pt,belowskip=-1pt}
    \begin{subfigure}[t]{\columnwidth}
        \includegraphics[width=\columnwidth]{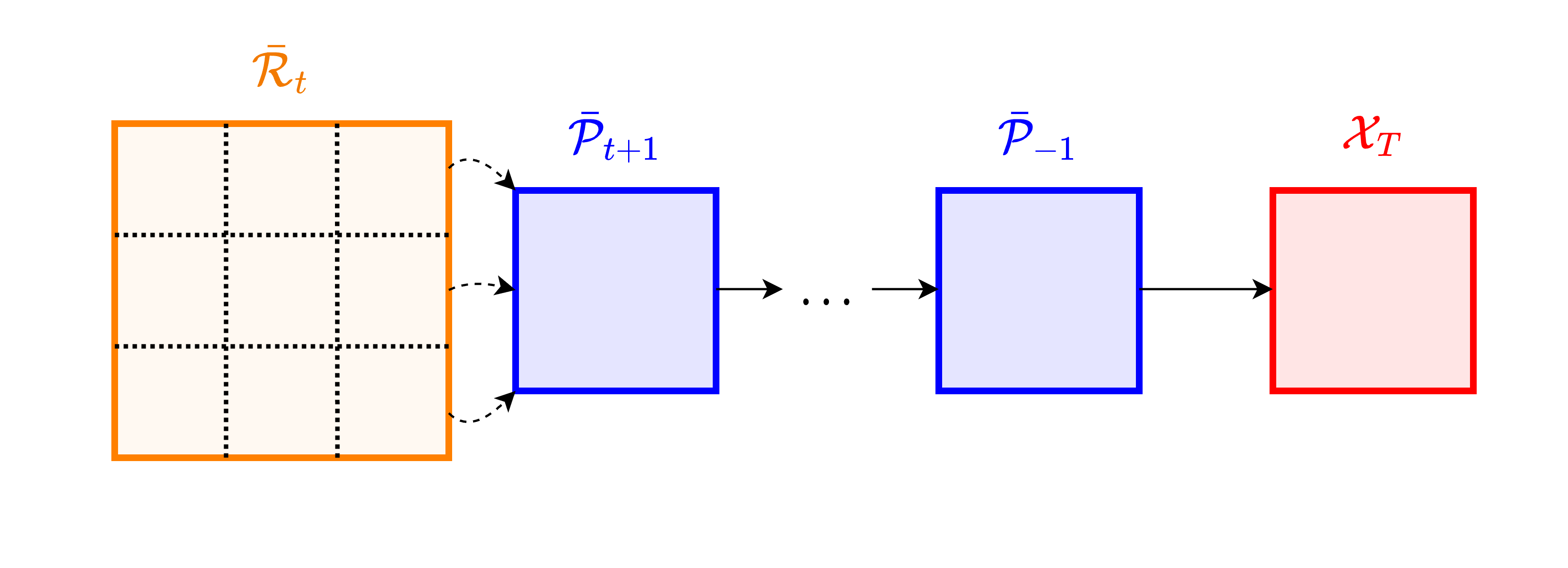}
\vspace*{-8mm}
        \caption{BRSP: At each step, $\bar{\mathcal{R}}_t$ is split and $\pi$ is relaxed over each partition to determine a set of states that reaches $\bar{\mathcal{P}}_{t+1},\ \bar{\mathcal{P}}_{t+2},\ ...,\ \mathcal{X}_T$.}
        \label{fig:approach:brsp_cartoon}
    \end{subfigure}
    \begin{subfigure}[t]{\columnwidth}
        \includegraphics[width=\columnwidth]{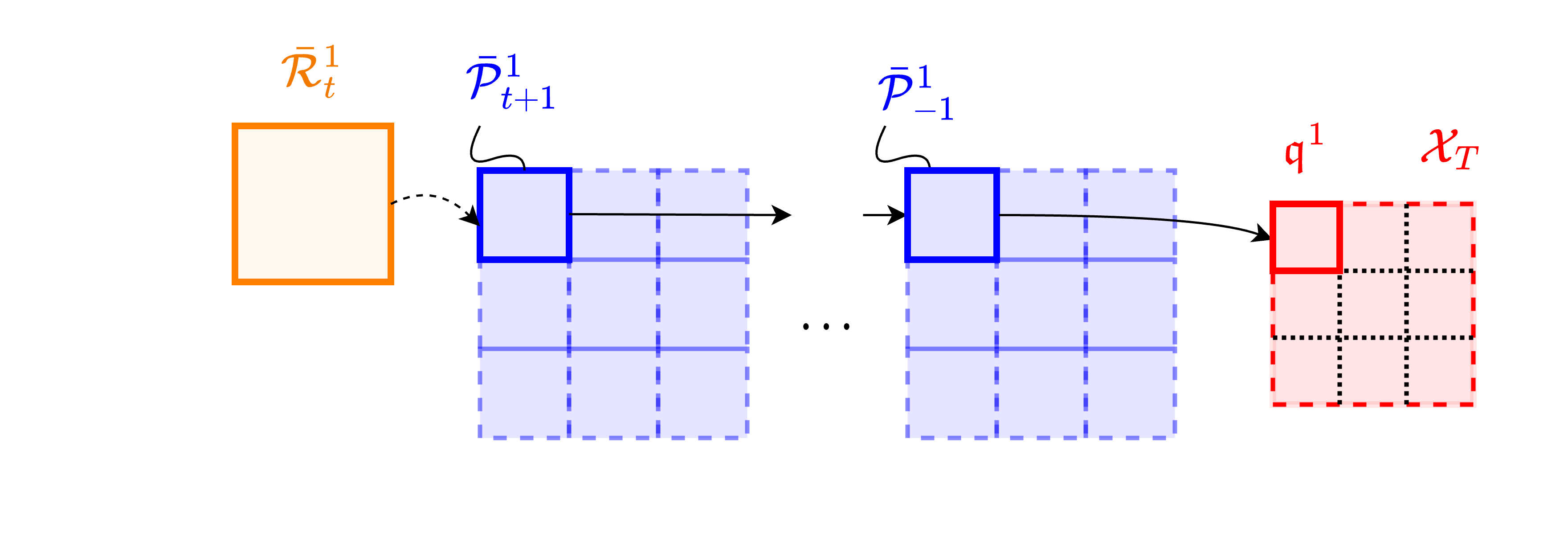}
        \vspace*{-8mm}
\caption{TSP: $\mathcal{X}_T$ is split and $\pi$ is relaxed over the entire BR set corresponding to the target set partition. each partition to determine a set of states that reaches $\bar{\mathcal{P}}_{t+1},\ \bar{\mathcal{P}}_{t+2},\ ...,\ \mathcal{X}_T$.}
        \label{fig:approach:tsp_cartoon}
    \end{subfigure}
    \label{fig:reach_comp}
    \begin{subfigure}[t]{\columnwidth}
        \includegraphics[width=\columnwidth]{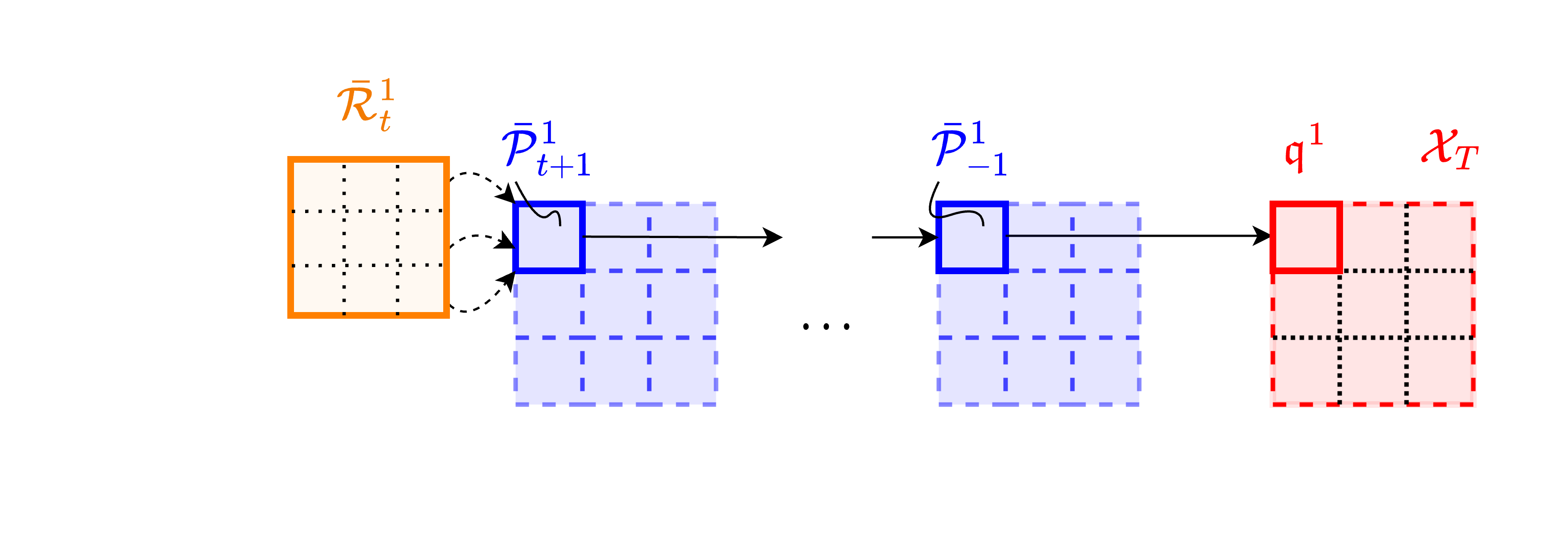}
\vspace*{-8mm}
        \caption{Hybrid Partitioning: TSP splits the problem into smaller parts, leading to smaller upstream BPOAs, and BRSP splits the BR set, giving tighter NN relaxations that are used to calculate the BPOA at time $t$.}
        \label{fig:approach:hybrid_cartoon}
    \end{subfigure}
    \caption{Summary of different partitioning strategies to reduce conservativeness in $\bar{\mathcal{P}}_t$.}
    \label{fig:approach:partitioning_cartoon}
\end{figure}

\subsection{BRSP,  TSP, \& Hybrid Partitioning} \label{sec:approach:tsp_vs_brsp}
A comparison of BRSP and TSP is shown in \cref{fig:approach:partitioning_cartoon}.
As explained in \cref{sec:prelim:partitioning} and depicted in \cref{fig:approach:brsp_cartoon}, BRSP is designed to directly obtain a tighter NN relaxation for time $t$ by splitting $\bar{\mathcal{R}}_t$.
Conversely, TSP, shown in \cref{fig:approach:tsp_cartoon}, splits $\mathcal{X}_T$ into smaller parts, each of which can be individually analyzed.
While TSP can reduce conservativeness because the BR sets of the target set partitions are smaller than that of the original target set (thus leading to a tighter relaxation than no partitioning at all), the BR sets can still be large, leading to a loose NN relaxation and a poor result.
However, TSP offers a potential advantage because it ultimately has to bound a subset of the true BP set (the BP corresponding to a particular region of $\mathcal{X}_T$), meaning that the BPOA can be smaller at each step, thereby allowing upstream NN relaxations to be tighter.
Thus, while BRSP was originally chosen over TSP in \cite{rober2022backward} because of its direct relationship with the NN relaxation, TSP unlocks the possibility for the tightened relaxations to be preserved over the time horizon.
This leads us to the conclusion that a hybrid partitioning approach, shown in \cref{fig:approach:hybrid_cartoon}, can be used to reduce conservativeness.
The key idea is that BRSP is used at each timestep to ensure that the BPOA is tightened to the BP, and TSP is used to keep the BPs small, allowing for tighter NN relaxations for the upstream BPOAs.
The net result is that the BPOAs are less conservative than is possible with BRSP or TSP alone.

\subsection{BPOA Calculation}

Keeping in mind the considerations discussed in \cref{sec:approach:tsp_vs_brsp}, we extend the method described in \cite{rober2022partition} with the addition of a TSP step.
Our approach is summarized in \cref{alg:hybridBackprojection} and detailed below.
The first step to our approach is to split $\mathcal{X}_T$ into a set of partitions $\mathcal{Q}$.
This TSP step can be done in several ways: in this paper we consider uniform partitioning where each dimension is split uniformly a predefined number of times, but more advanced strategies analogous to those introduced in \cite{everett2020robustness, rubies2019fast} could also be used.
Next, we must calculate $\bar{\mathcal{P}}^i_t$ for each $i^\mathrm{th}$ partition element $\mathfrak{q}_i \in \mathcal{Q}$.
To calculate $\bar{\mathcal{P}}^i_t$, we first calculate $\bar{\mathcal{R}}^i_t$ by solving
\begin{equation}
\bar{\bar{\mathbf{x}}}_{t; \istate} = \min_{\mathbf{x}_t, \mathbf{u}_t \in \mathcal{F}_{\bar{\mathcal{R}}_t}} \mathbf{e}_k^\top  \mathbf{x}_{t}, \quad \quad \ubar{\ubar{\mathbf{x}}}_{t; \istate} = \max_{\mathbf{x}_t, \mathbf{u}_t \in \mathcal{F}_{\bar{\mathcal{R}}_t}} \mathbf{e}_k^\top  \mathbf{x}_{t}, \label{eqn:backreachable:lp_obj}
\end{equation}
for each state $\istate \in [n_x]$, where $\mathbf{x}_{t;k}$ denotes the $k^\mathrm{th}$ element of $\mathbf{x}_t$ and the set of constraints $\mathcal{F}_{\bar{\mathcal{R}}_t}$ is defined as
\begin{equation}
\mathcal{F}_{\bar{\mathcal{R}}_t} \triangleq \left\{
\mathbf{x}_t, \mathbf{u}_t\ \left| \
\begin{aligned}
& \mathbf{A} \mathbf{x}_t + \mathbf{B} \mathbf{u}_t + \mathbf{c} \in \bar{\mathcal{P}}^i_{t+1},\\
& \mathbf{u}_t \in \mathcal{U}, \\ 
& \mathbf{x}_t \in \mathcal{X}, \\ 
\end{aligned}\right.\kern-\nulldelimiterspace
\right\},
\label{eqn:backreachable:lp_constr}
\end{equation}
where $\bar{\mathcal{P}}^i_{0} \triangleq \mathfrak{q}_i$ and where $\mathbf{e}_k \ \in \R^{n_x}$ denotes the indicator vector with $\mathbf{e}_{k;k} = 1$ and $\mathbf{e}_{k;\ell} = 0\ \forall \ell \in [n_x] \setminus k$. $\bar{\mathcal{R}}^i_{t}$ is then constructed by setting $\bar{\mathcal{R}}^i_{t} \triangleq \{\mathbf{x}\ | \ \ubar{\ubar{\mathbf{x}}}_{t} \leq \mathbf{x} \leq \bar{\bar{\mathbf{x}}}_{t}\}$.

The set $\bar{\mathcal{R}}^i_t$ provides a region over which the NN can be relaxed using \cite{zhang2018efficient}. However, as discussed in \cref{sec:approach:tsp_vs_brsp} and \cite{rober2022partition}, relaxing the NN over all of $\bar{\mathcal{R}}^i_t$ (even with TSP) can cause conservative bounds, thus we implement BRSP to get the set of BR set partitions $\mathcal{S}$. For each $j^\mathrm{th}$ partition element $\mathfrak{s}_j \in \mathcal{S}$ we use \cite{zhang2018efficient} to obtain $\pi^{L^i_j}_t$ and $\pi^{U^i_j}_t$, thus allowing us to construct the constraint set
\begin{equation}
\mathcal{F}_{\bar{\mathcal{P}}_t} \! \triangleq \! \left\{
\! \mathbf{x}_{\mathfrak{t}}, \mathbf{u}_{\mathfrak{t}} \left| \ 
\begin{aligned}
& \x_\mathfrak{t} \in \bar{\mathcal{R}}_\mathfrak{t} \\
& \mathbf{u}_\mathfrak{t} \in [\pi^L_\mathfrak{t}(\x_\mathfrak{t}), \pi^U_\mathfrak{t}(\x_\mathfrak{t})] \\
& \x_{\mathfrak{t}+1} = f(\mathbf{x}_\mathfrak{t}, \mathbf{u}_\mathfrak{t}) \\
& \x_{\mathfrak{t}+1} \in \bar{\mathcal{P}}_{\mathfrak{t}+1} \\ 
\end{aligned} \; \; \forall \mathfrak{t}\in \{t, ..., -1\} \right.\kern-\nulldelimiterspace 
\right\},
\label{eqn:hybrid:lp_constr}
\end{equation}
where 
\begin{equation}
    \begin{cases}
        \begin{aligned}
            (\pi^L_\mathfrak{t},\ \pi^U_\mathfrak{t}) \leftarrow (\pi^{L^i_j}_t,\ \pi^{U^i_j}_t),
        \end{aligned}
        \quad \mathfrak{t} = t \\ \\
        \begin{aligned}
            (\pi^L_\mathfrak{t},\ \pi^U_\mathfrak{t}) \leftarrow (\bar{\pi}^{L^i}_\mathfrak{t},\ \bar{\pi}^{U^i}_\mathfrak{t})
        \end{aligned},
        \quad \mathfrak{t} \in \{t+1, ..., -1\}
    \end{cases}
\end{equation}
where $(\bar{\pi}^{L^i}_t,\ \bar{\pi}^{U^i}_t)$ are affine bounds valid over $\bar{\mathcal{P}}^i_{\mathfrak{t}}$.
The constraint set defined in \cref{eqn:hybrid:lp_constr} is then used to solve 
\begin{equation}
\begin{split}
& \bar{\mathbf{x}}_{t; \istate} = \min_{\mathbf{x}_{t:-1}, \mathbf{u}_{t:-1} \in \mathcal{F}_{\bar{\mathcal{P}}_t}} \mathbf{e}_k^\top \mathbf{x}_{t}, \\ & \ubar{\mathbf{x}}_{t; \istate} = \max_{\mathbf{x}_{t:-1}, \mathbf{u}_{t:-1} \in \mathcal{F}_{\bar{\mathcal{P}}_t}} \mathbf{e}_k^\top \mathbf{x}_{t}. \label{eqn:hybrid:lp_obj}
\end{split}
\end{equation}
for each state $\istate \in [n_x]$.
From this we obtain $\bar{\mathcal{P}}^i_{t} \triangleq \{\mathbf{x}\ | \ \ubar{\mathbf{x}}_{t} \leq \mathbf{x} \leq \bar{\mathbf{x}}_{t}\}$ and relax $\pi$ over $\bar{\mathcal{P}}^i_{t}$ to get $\bar{\pi}^{L^i}_t$ and $\bar{\pi}^{U^i}_t$ for use during the calculation of downstream BPOAs.
Finally, once $\bar{\mathcal{P}}^i_{t}$ has been calculated for all $\mathfrak{q}_i \in \mathcal{Q}$ and for all $t \in \mathcal{T}$, we construct our final set of BPOAs over the time horizon $\tau$ via 
\begin{equation}
    \bar{\mathcal{P}}_{t} = \bigcup_{\mathfrak{q}_i \in \mathcal{Q}} \bar{\mathcal{P}}^i_{t}\quad \forall t \in \mathcal{T}.
\end{equation}
Note that the number of LPs solved during the execution of \cref{alg:hybridBackprojection} is $N_{LP} = 2 n_x |\mathcal{S}| |\mathcal{Q}| \tau$, which has an additional term associated with TSP.
However, as is shown in \cref{sec:results}, the increase associated with $|\mathcal{Q}|$ can be matched with a reduction in $|\mathcal{S}|$, resulting in lower error in the same computation time.

\begin{algorithm}[t]
 \caption{\hybrid{}}
 \begin{algorithmic}[1]
 \setcounter{ALC@unique}{0}
 \renewcommand{\algorithmicrequire}{\textbf{Input:}}
 \renewcommand{\algorithmicensure}{\textbf{Output:}}
 \REQUIRE target state set $\mathcal{X}_T$, trained NN control policy $\pi$, time horizon $\tau$, partition parameters $(r_1, r_2)$, minimum partition element volume $v_m$
 \ENSURE BP set approximations $\bar{\mathcal{P}}_{-\tau:0}$
    \STATE $\mathcal{Q} \leftarrow \mathrm{TSP}(\mathcal{X}_T, r_1)$
    \STATE $\bar{\mathcal{P}}_{-\tau:0} \leftarrow \emptyset$
    \FOR{$\mathfrak{q}$ in $\mathcal{Q}$}
        \STATE $\bar{\mathcal{P}}^\mathfrak{q}_{-\tau:0} \leftarrow \emptyset$
        \STATE $\mathbf{\Omega}^{\mathfrak{q}}_{-\tau:-1} \leftarrow \emptyset$
        \STATE $\bar{\mathcal{P}}^{\mathfrak{q}}_{0} \leftarrow \mathfrak{q}$
        \FOR{$t$ in $\{-1, \ldots, -\tau\}$} \label{alg:hybridBackprojection:timestep_for_loop}
            \STATE $\bar{\mathcal{R}}_{t}\! \leftarrow\! \mathrm{backreach}(\bar{\mathcal{P}}^{\mathfrak{q}}_{t\text{+}1}, \mathcal{U}, \mathcal{X})$
            \STATE $\mathcal{S} \leftarrow \mathrm{BRSP}(\bar{\mathcal{R}}_{t}, \mathfrak{q}, \pi, t, r_2, v_m)$
            \FOR{$\mathfrak{s}$ in $\mathcal{S}$}
                \STATE $[\ubar{\ubar{\mathbf{x}}}_{t}, \bar{\bar{\mathbf{x}}}_{t}] \leftarrow \mathfrak{s}$
                \STATE $\pi^{L^{\mathfrak{q}}_{\mathfrak{s}}}, \pi^{U^{\mathfrak{q}}_{\mathfrak{s}}} \leftarrow \mathrm{CROWN}(\pi, [\ubar{\ubar{\mathbf{x}}}_{t}, \bar{\bar{\mathbf{x}}}_{t}])$
                \FOR{$\istate \in n_x$}
                    \STATE $\bar{\mathbf{x}}_{t;k} \leftarrow \! \mathrm{LpMax}(\bar{\mathcal{P}}^{\mathfrak{q}}_{t:0}, \mathbf{\Omega}^{\mathfrak{q}}_{t:\text{-}1},\pi^{L^{\mathfrak{q}}_{\mathfrak{s}}}, \pi^{U^{\mathfrak{q}}_{\mathfrak{s}}})$
                    \STATE $\ubar{\mathbf{x}}_{t;k} \leftarrow \! \mathrm{LpMin}(\bar{\mathcal{P}}^{\mathfrak{q}}_{t:0},\ \mathbf{\Omega}^{\mathfrak{q}}_{t:\text{-}1},\pi^{L^{\mathfrak{q}}_{\mathfrak{s}}}, \pi^{U^{\mathfrak{q}}_{\mathfrak{s}}})$
                \ENDFOR
                \STATE $\mathcal{A} \leftarrow \{\mathbf{x}\ \lvert\ \forall \istate \in n_x,\ \ubar{\mathbf{x}}_{t;k} \leq \mathbf{x} \leq \bar{\mathbf{x}}_{t;k}\}$
                \STATE $\bar{\mathcal{P}}^{\mathfrak{q}}_{t} \leftarrow \bar{\mathcal{P}}^{\mathfrak{q}}_{t} \cup \mathcal{A}$ \label{alg:hybridBackprojection:union}
                \STATE $\bar{\mathcal{P}}^{\mathfrak{q}}_{t} = [\ubar{\mathbf{x}}'_{t}, \bar{\mathbf{x}}'_{t}] \leftarrow \text{boundRectangle}(\bar{\mathcal{P}}^{\mathfrak{q}}_{t})$ 
                \label{alg:hybridBackprojection:bound}
            \ENDFOR
            \STATE $\bar{\mathcal{P}}_{t} \leftarrow \bar{\mathcal{P}}^{\mathfrak{q}}_{t} \cup \bar{\mathcal{P}}_{t}$
            \STATE $\mathbf{\Omega}^{\mathfrak{q}}_{t} = [\bar{\pi}^{L^{\mathfrak{q}}}, \bar{\pi}^{U^{\mathfrak{q}}}] \leftarrow \mathrm{CROWN}(\pi, [\ubar{\mathbf{x}}'_{t}, \bar{\mathbf{x}}'_{t}])$
        \ENDFOR
    \ENDFOR
 \RETURN $\bar{\mathcal{P}}_{-\tau:0}$ \label{alg:hybridBackprojection:return}
 \end{algorithmic}\label{alg:hybridBackprojection}
\end{algorithm}



\section{Numerical Results}
\label{sec:results}

In this section, we make use of a double integrator model with dynamics given by 
\begin{equation}
    \mathbf{x}_{t+1} =
    \underbrace{
    \begin{bmatrix}
    1 & 1 \\
    0 & 1
    \end{bmatrix}}_{\mathbf{A}} \mathbf{x}_t +
    \underbrace{
    \begin{bmatrix}
    0.5 \\ 1
    \end{bmatrix}}_{\mathbf{B}} \mathbf{u}_t
\end{equation}
with discrete sampling time $\Delta t=1$s to investigate our approach and validate the ideas discussed in \cref{sec:approach}.
The NN controller from \cite{everett2021reachability} has $[5,5]$ neurons, ReLU activations and was trained with state-action pairs generated by an MPC controller.


\begin{figure}[t]
\setlength\belowcaptionskip{-0.7\baselineskip}
\centering
\captionsetup[subfigure]{aboveskip=-1pt,belowskip=-1pt}
    \begin{subfigure}[t]{0.9\columnwidth}
        \begin{tikzpicture}[fill=white]
            \node[anchor=south west,inner sep=0] (image) at (0,0) {\includegraphics[width=\columnwidth]{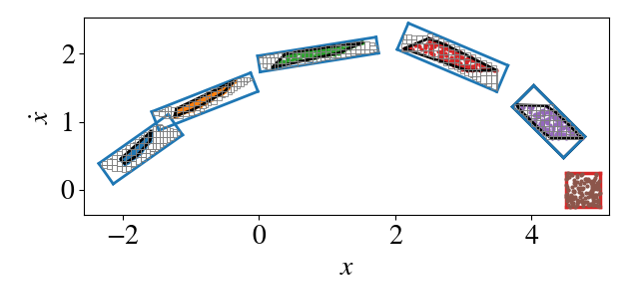}};
        \end{tikzpicture}
        \caption{Backward reachability analysis using rotated-rectangle set representation.}
        \label{fig:results:set_representation:rotated_rectangles}
    \end{subfigure}
    \begin{subfigure}[t]{0.9\columnwidth}
        \begin{tikzpicture}[fill=white]
            \node[anchor=south west,inner sep=0] (image) at (0,0) {\includegraphics[width=\columnwidth]{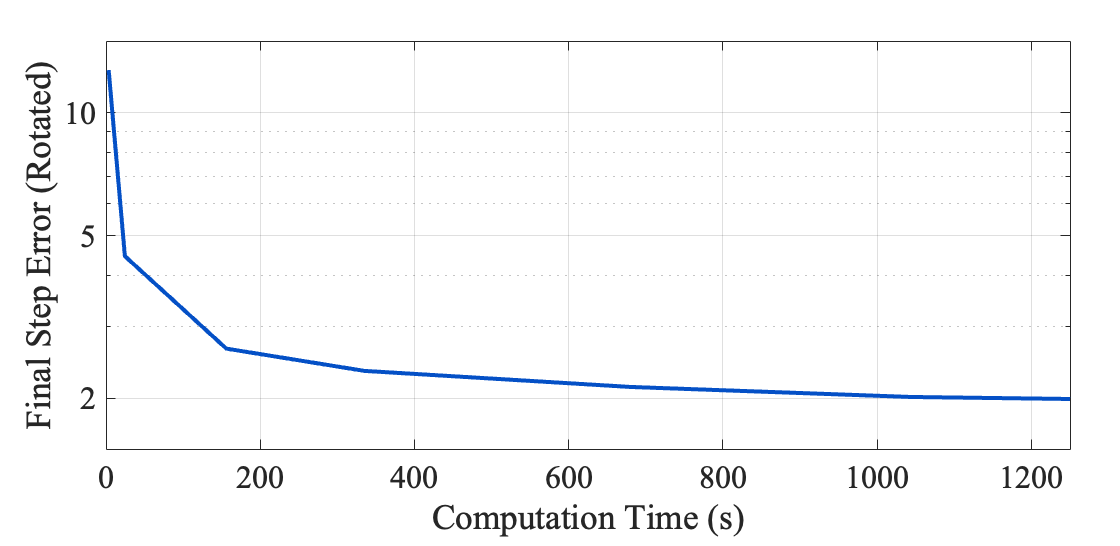}};
        \end{tikzpicture}
        \caption{Error vs computation time using rotated-rectangle set representation with only BRSP, demonstrating that the existence of a lower error bound is not specific to an axis-aligned hyper-rectangular set representation.}
        \label{fig:results:set_representation:error}
    \end{subfigure}
    \caption{Lower bound on error still exists with a better set representation.}
    \label{fig:results:set_representation}
\end{figure}

First, we address the idea that an alternative solution to the problem described in \cref{fig:prelim:pure_brsp_limit} could be to use a better set representation than the hyper-rectangles used in \cite{rober2022partition}.
While more complex set representations could indeed offer a tighter approximation to the true BP sets, they still run into the problem of a lower bound on the approximation error.
To demonstrate this point, \cref{fig:results:set_representation} shows BRSP with 100 guided BR partitions where the BPOAs are formulated as rotated hyper-rectangular bounds (\cref{fig:results:set_representation:rotated_rectangles}, obtained by replacing \cref{alg:hybridBackprojection:bound} of \cref{alg:hybridBackprojection} with a function that finds the minimum-area rotated rectangle to bound the BR partitions).
The approximation error shown in \cref{fig:results:set_representation:error} represents the error in the final BPOA, calculated as
\begin{equation}
    \mathrm{error} = \frac{A_{\mathrm{BPOA}} - A_\mathrm{true}}{A_\mathrm{true}},
    \label{eqn:BP_estimate_error}
\end{equation}
where $A_\mathrm{true}$ denotes the area of the tightest rotated-rectangular bound of the true BP set, calculated using Monte Carlo simulations, and $A_\mathrm{BPOA}$ denotes the area of the BPOA.
Given that there exist a point where further BRSP does not tighten the BPOAs as shown in \cref{fig:results:set_representation:error}, we can see that the rotated rectangles exhibit the problem described in \cref{fig:prelim:pure_brsp_limit}, i.e., an imperfect set representation coupled with inexact NN relaxations results in a lower bound to the approximation error.

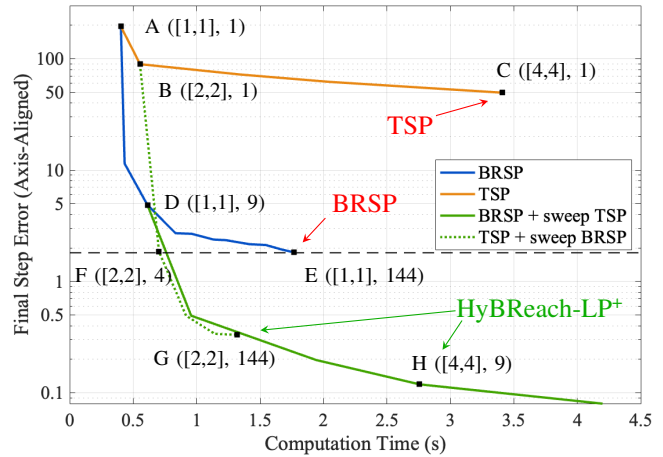
\begin{figure}[t]
\centering
    \begin{tikzpicture}[fill=white]
        \node[anchor=south west,inner sep=0] (image) at (0,0) {\includegraphics[width=\columnwidth]{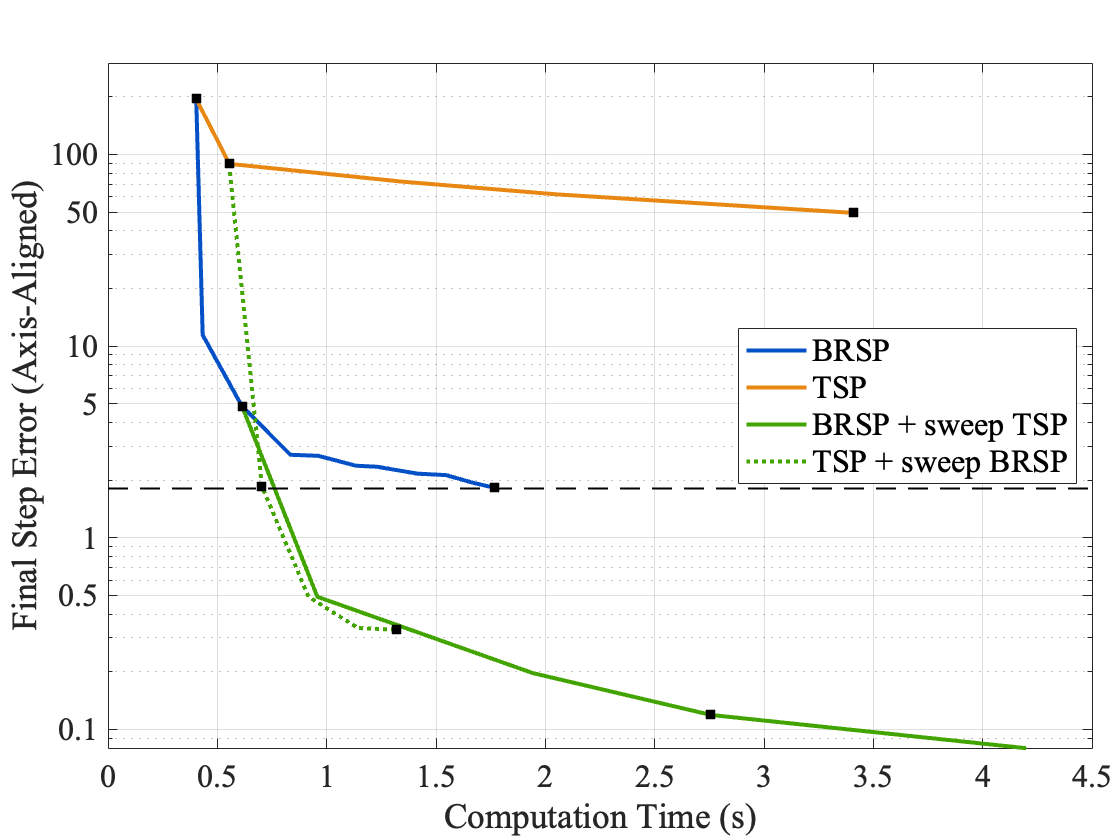}};
        \begin{scope}[x={(image.south east)},y={(image.north west)}]
            \node[] at (0.29,0.88) {\footnotesize A ([1,1], 1)};
            \node[] at (0.83,0.79) {\footnotesize C ([4,4], 1)};
            \node[] at (0.55,0.37) {\footnotesize E ([1,1], 144)};
            \node[] at (0.32,0.52) {\footnotesize D ([1,1], 9)};
            \node[] at (0.18,0.37) {\footnotesize F ([2,2], 4)};
            \node[] at (0.31,0.75) {\footnotesize B ([2,2], 1)};
            \node[] at (0.32,0.20) {\footnotesize G ([2,2], 144)};
            \node[] at (0.7,0.19) {\footnotesize H ([4,4], 9)};
            \node[] at (0.55,0.52) {\color{red} BRSP};
            \draw [red, -stealth](0.49,0.49) -- (0.455,0.44);
            \node[] at (0.62,0.68) {\color{red} TSP};
            \draw [red, -stealth](0.66,0.7) -- (0.74,0.73);
            
            \definecolor{nice_green}{HTML}{0BAD01}
            \node[nice_green] at (0.82,0.3){{\hybrid{}}};
            \draw [nice_green, -stealth](0.7,0.28) -- (0.67,0.22);
            \draw [nice_green, -stealth](0.69,0.3) -- (0.39,0.26);
        \end{scope}
    \end{tikzpicture}
    \caption{\hybrid{} (green) reduces error more quickly than BRSP or TSP alone and drops below the lower error bound for BRSP (dashed black). As BRSP is increased from Point B (dotted green), a lower-error bound is encountered at Point G, indicating error cannot be reduced for [2,2] TSP. As TSP in increased from Point D, error continues to decrease without an apparent bound.}
    \label{fig:results:hybrid_vs_brsp_vs_tsp}
\end{figure}

Next, in \cref{fig:results:hybrid_vs_brsp_vs_tsp}, we show how hybrid partitioning performs better than BRSP or TSP alone by plotting the approximation error vs. calculation time using different partitioning configurations.
Specifically we show the error for the final step in a 5 s time horizon averaged over 5 target state sets randomly distributed over the positive x-axis.
The partitioning configurations of several points (A-H) are highlighted using the notation 
\begin{equation*}
    \mathbf{r} = (\mathrm{TSP},\ \mathrm{BRSP}) = ([n_T,n_T],\ n_B),
\end{equation*}
where $[n_T,n_T]$ indicates TSP is performed with a uniform grid of $n_T \times n_T$ partitions, and BRSP has a budget of $n_B$ guided partitions \cite{rober2022partition}.
Note that for \cref{fig:results:hybrid_vs_brsp_vs_tsp}, $A_\mathrm{true}$ from \cref{eqn:BP_estimate_error} denotes the area of the tightest \textit{axis-aligned} rectangle, meaning that zero approximation error is theoretically possible.
Note also that the difference in timescale between \cref{fig:results:set_representation} and \cref{fig:results:hybrid_vs_brsp_vs_tsp} can be attributed to the fact that the BRSP algorithm in \cite{rober2022partition} was designed to efficiently calculate axis-aligned hyper-rectangular bounds, and some of the computational improvements did not transfer to the current implementation of the rotated rectangles.

\cref{fig:results:hybrid_vs_brsp_vs_tsp} shows that given the options of BRSP, TSP and hybrid partitioning (\hybrid{}), TSP (orange) performs the worst. This is not surprising because BR sets can be large, leading to loose NN relaxations even for relatively fine TSP.
BRSP performs better, but as discussed previously, encounters a lower bound (dashed black) on the error that cannot be crossed without a more advanced partitioning strategy.
Finally, hybrid partitioning (solid and dotted green) is shown to perform better than both BRSP and TSP.
Starting from Point B ($[2,2]$ TSP), error drops sharply as the BR set partitions are increased until Point G, when again there is a lower error bound that cannot be crossed without further TSP.
Similarly, starting from Point D (BRSP with 9 guided partitions \cite{rober2022partition}), error drops sharply and continues to decrease to Point H and beyond.
From \cref{fig:results:hybrid_vs_brsp_vs_tsp}, we can see that not only does the hybrid partitioning strategy allow the approximation error to drop below the bound of what is possible with BRSP, but it does so in the same computation time.
This highlights the idea that although hybrid partitioning breaks the problem into more elements ($|\mathcal{Q}| * |\mathcal{S}|$), thereby increasing the expected computational load, by distributing the same computational load between TSP and BRSP, lower error can be achieved without sacrificing computation time.

\begin{figure}[t]
    \begin{subfigure}[t]{\columnwidth}
        \centering
        \begin{tikzpicture}[fill=white]
            \node[anchor=south west,inner sep=0] (image) at (-0.5,0) {\includegraphics[width=0.54\columnwidth]{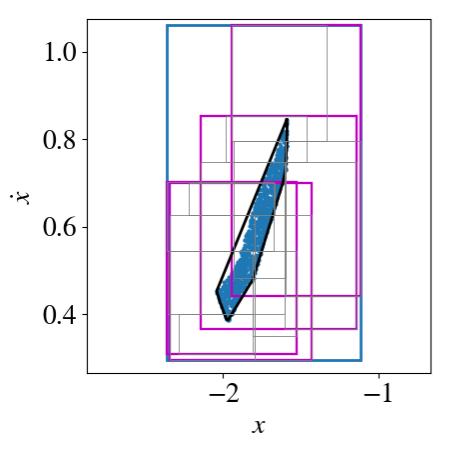}};
            \node[anchor=south west,inner sep=0] (image) at (4.0,0.82) {\includegraphics[width=0.36\columnwidth]{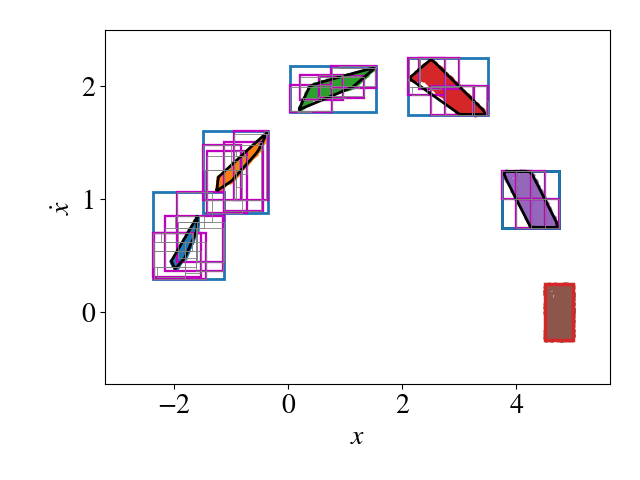}};
            \draw [draw=black,dashed,line width=0.2mm] (4.66,1.7) rectangle ++(0.55,0.53);
            \draw[line width=0.2mm] (4.66,2.23) -- (3.8,4.32);
            \draw[line width=0.2mm] (4.66,1.7) -- (3.8,0.78);
        \end{tikzpicture}
\vspace*{-2mm}
        \caption{Hybrid partitioning with 4 guided BR set partitions and [2,2] uniform target set partitions (Point D in \cref{fig:results:hybrid_vs_brsp_vs_tsp}) showing $\bar{\mathcal{P}}^{1:4}_{-5}$ (magenta) and their BR set partitions (grey).}
        \label{fig:results:selected_partitioning_examples:a}
    \end{subfigure}
    \begin{subfigure}[t]{\columnwidth}
        \centering
        \begin{tikzpicture}[fill=white]
            \node[anchor=south west,inner sep=0] (image) at (-0.5,0) {\includegraphics[width=0.54\columnwidth]{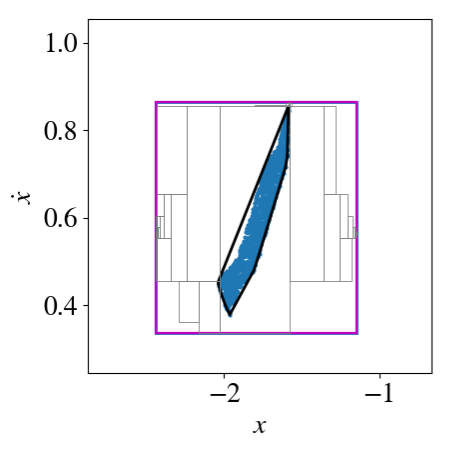}};
            \node[anchor=south west,inner sep=0] (image) at (4.0,0.82) {\includegraphics[width=0.36\columnwidth]{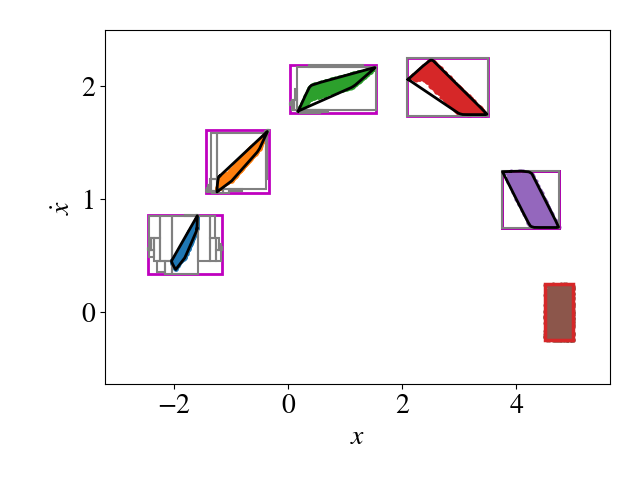}};
            \draw [draw=black,dashed,line width=0.2mm] (4.66,1.7) rectangle ++(0.55,0.53);
            \draw[line width=0.2mm] (4.66,2.23) -- (3.8,4.32);
            \draw[line width=0.2mm] (4.66,1.7) -- (3.8,0.78);
        \end{tikzpicture}
\vspace*{-2mm}
        \caption{BRSP with 144 BR set partitions (Point E in \cref{fig:results:hybrid_vs_brsp_vs_tsp}). Further BRSP cannot reduce error.}
        \label{fig:results:selected_partitioning_examples:b}
    \end{subfigure}
    \begin{subfigure}[t]{\columnwidth}
        \centering
        \begin{tikzpicture}[fill=white]
            \node[anchor=south west,inner sep=0] (image) at (-0.5,0) {\includegraphics[width=0.54\columnwidth]{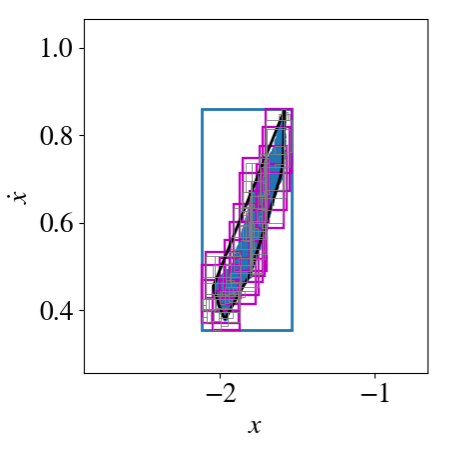}};
            \node[anchor=south west,inner sep=0] (image) at (4.0,0.82) {\includegraphics[width=0.36\columnwidth]{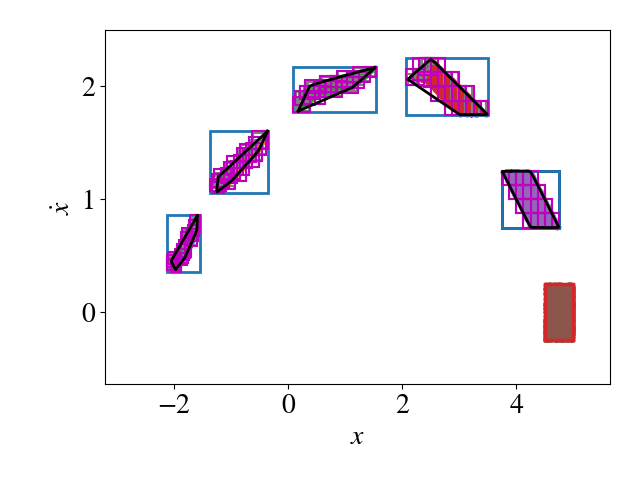}};
            \draw [draw=black,dashed,line width=0.2mm] (4.66,1.7) rectangle ++(0.55,0.53);
            \draw[line width=0.2mm] (4.66,2.25) -- (3.75,4.32);
            \draw[line width=0.2mm] (4.66,1.7) -- (3.75,0.78);
        \end{tikzpicture}
\vspace*{-2mm}
        \caption{Hybrid partitioning with 9 guided BR set partitions and [4,4] uniform target set partitions (Point H in \cref{fig:results:hybrid_vs_brsp_vs_tsp}). Approximation error approaches zero.}
        \label{fig:results:selected_partitioning_examples:c}
    \end{subfigure}
    \caption{Hybrid partitioning reduces approximation error beyond what is possible with BRSP alone.}
    \label{fig:results:selected_partitioning_examples}
\vspace*{-5mm}
\end{figure}
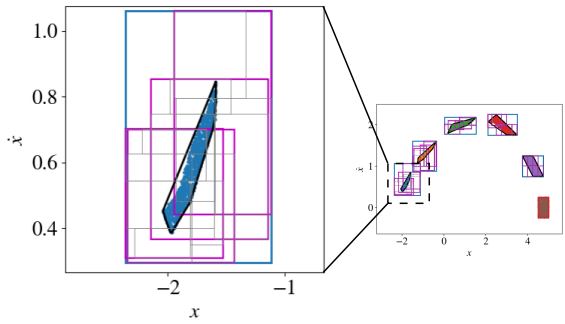
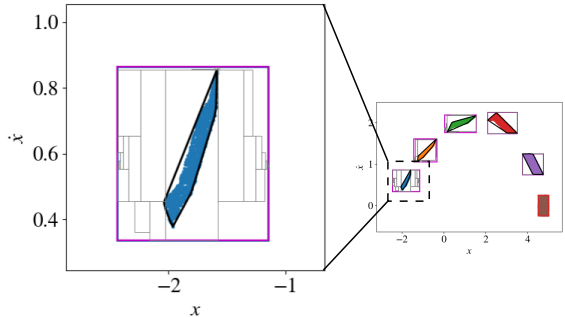
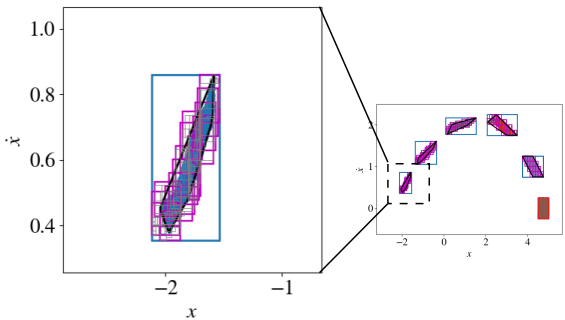

\cref{fig:results:selected_partitioning_examples} shows three points selected from \cref{fig:results:hybrid_vs_brsp_vs_tsp} to visualize some partitioning configurations of interest.
\cref{fig:results:selected_partitioning_examples:a} shows Point D ([2,2], 4), a simple example of hybrid partitioning.
Notice the 4 BPOA components ($\mathcal{P}^{1:4}_{-5}$, magenta) associated with the target set partitions, each with 4 BR set partitions (grey) within them.
\cref{fig:results:selected_partitioning_examples:b} shows Point E ([1,1], 144).
Without any TSP, the lone magenta BPOA component is identical to the aggregate BP $\bar{\mathcal{P}}_{-5}$ and the grey BR partitions have converged to the lowest possible error.
Notice that as discussed in \cite{rober2022partition}, the guided partitioning method is designed to shrink the axis-aligned bounding box efficiently, which explains the fine partitioning towards the edges of the BPOA; no partitioning within the inner elements can change the error value because they would not affect any extreme value within $\bar{\mathcal{P}}_{-5}$.
Finally, \cref{fig:results:selected_partitioning_examples:c} shows Point H ([4,4], 9).
Notice that the BPOA bounds the true BP set much more tightly than any other case.

\section{Conclusion}

This paper considered the problem of backward reachability for neural feedback loops.
Specifically, it sought to improve upon previous work by reducing conservativeness in the calculation of BPOAs.
Reduced conservativeness was accomplished via the introduction of a novel hybrid partitioning algorithm \hybrid{} that combines the concepts of TSP and BRSP.
Numerical experiments were conducted to validate the approach, demonstrating a near-order of magnitude reduction in approximation error, given the same computation time as TSP and BRSP.

\balance
\bibliographystyle{IEEEtran}
\bibliography{refs}

\end{document}